\documentclass[fleqn,usenatbib]{mnras}

\usepackage[T1]{fontenc}

\DeclareRobustCommand{\VAN}[3]{#2}
\let\VANthebibliography\thebibliography
\def\thebibliography{\DeclareRobustCommand{\VAN}[3]{##3}\VANthebibliography}

\usepackage{graphicx}	
\usepackage{amsmath}	
\usepackage{amssymb}	

\title[Multiphase Outflows in Cygnus A]{Powerful multiphase outflows in the central region of Cygnus A}

\author[R. A. Riffel]{
R. A. Riffel,$^{1}$\thanks{E-mail: rogemar@ufsm.br}\\
$^{1}$Universidade Federal de Santa Maria, Av. Roraima 1000, Cep 97105-900, Santa Maria, Brazil\\
}

\date{Accepted XXX. Received YYY; in original form ZZZ}

\pubyear{2021}

\begin{document}
\label{firstpage}
\pagerange{\pageref{firstpage}--\pageref{lastpage}}
\maketitle

\begin{abstract}
We use Gemini Near-Infrared Integral Field Spectrograph (NIFS) observations of the inner 3.5$\times$3.5 kpc$^2$ of the radio galaxy Cygnus A to map the gas excitation and kinematics at a spatial resolution of 200 pc. 
The emission of the ionised gas shows a biconical morphology, with half-opening angle of 45$^\circ$ and oriented along the position angle of the radio jet. Coronal line emission is seen within the cone, up to 1.75 kpc from the nucleus, with higher ionisation gas observed in the easterly side. The H$_2$ and [Fe\,{\sc ii}] emission lines are consistent with excitation by the central AGN, with some contribution of shocks to the southwest of the nucleus. The gas visual extinction and electron density are larger  than those from optical-based measurements, consistent with the fact that near-IR observations penetrate deeply into the gas emission structure, probing denser and more obscured regions. The gas kinematics shows two components: (i) a rotating disc with kinematic position angle of $\Psi_0=21^\circ\pm2^\circ$, seen both in  ionised and molecular gas, and (ii)  outflows with velocities of up to 600 km\,s$^{-1}$ observed within the ionisation cone in ionised gas and restricted to inner 0.5 arcsec in molecular gas.  The mass outflow rate in ionised gas is in the range $\sim100-280 {\rm M_\odot\, yr^{-1}}$ and the kinetic power of the outflow corresponds to 0.3--3.3 per cent of the AGN bolometric luminosity, indicating that the outflows in Cygnus A may be effective in suppressing star formation.
 \end{abstract}




\section{Introduction}

\begin{figure*}
\includegraphics[width=0.49\textwidth]{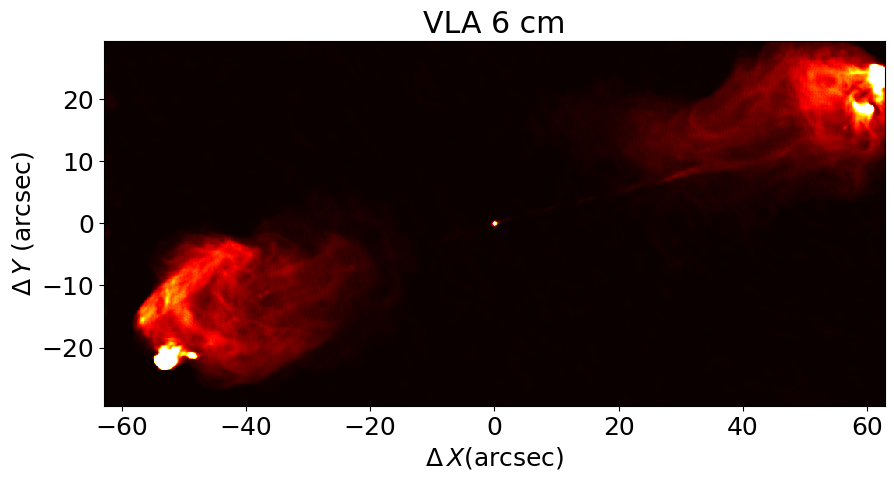}
\includegraphics[width=0.49\textwidth]{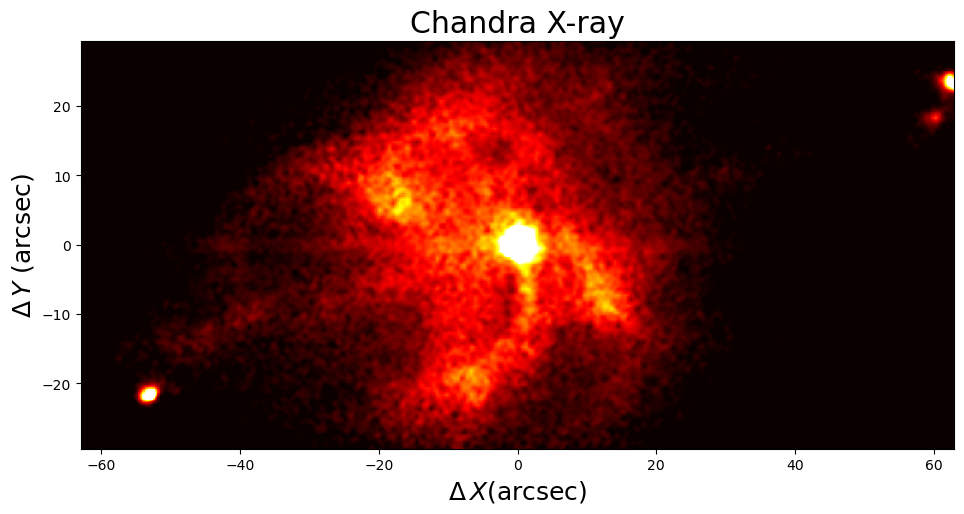}
\includegraphics[width=0.70\textwidth]{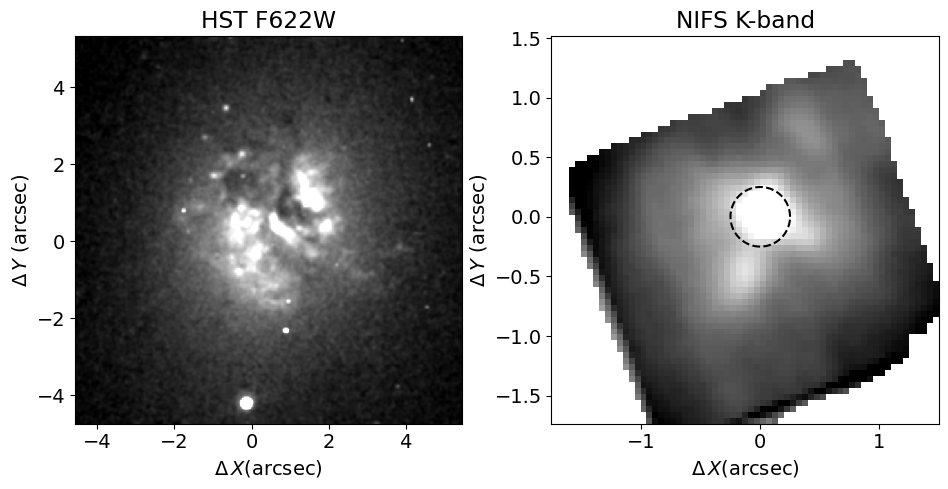}
\includegraphics[width=0.98\textwidth]{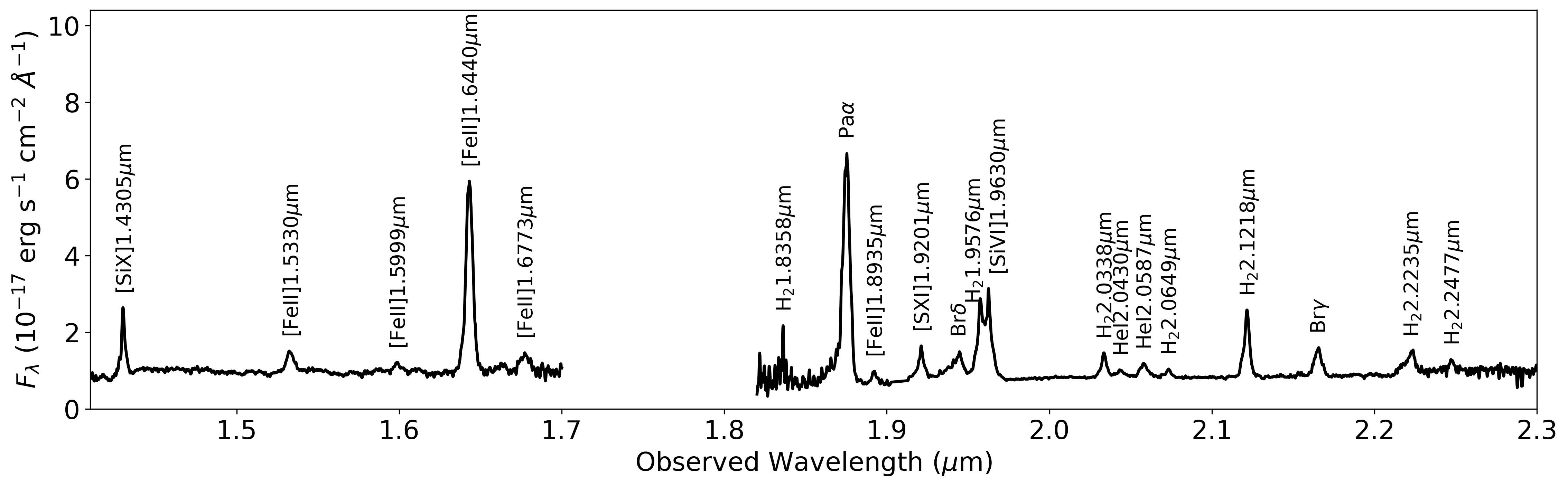}
\caption{Top: VLA 6 cm image of Cygnus A from \citet{perley98} and Chandra AGIS X-ray archival image (PI: M. Wise). Middle: The left panel shows the HST F622W archival image (Proposal ID: 5104; PI: J. Westphal) and the right panel shows the NIFS K-band continuum image, obtained as the average of the third order polynomial, used to reproduce continuum spectra, in the spectral range from 2.0--2.2 $\mu$m.   Bottom: H+K band nuclear spectrum of Cygnus\,A extracted within the circular aperture with 0\farcs25 radius, delineated by the circle on the K-band continuum image.}
\label{large}
\end{figure*}

Feedback from Active Galactic Nuclei (AGN) is a critical transformation mechanism of galaxies from star-forming to quiescent, coupling the growth of the central supermassive black holes (SMBHs)
and their host galaxies \citep[e.g.][]{cataneo09,alexander12,conselice14,harrison17} and being responsible for the correlation between the mass of the SMBH and the mass of the galaxy bulge \citep[e.g][]{magorian98,gebhardt00,ferrarese00,ferrasese05}.
AGN feedback is a strong function of luminosity and high-luminosity AGN inject enough energy into the surrounding medium so that the wind can overcome the inertia of the gas in the galactic potential \citep{harrison18}. Cygnus A is one of the most powerful  AGN in the local Universe and an ideal laboratory to investigate the effect of the AGN on the host galaxy, by mapping the gas kinematics and emission structure in the nuclear region in details.

Cygnus A is a narrow-line radio galaxy \citep{osterbrock75}, its redshift is $z=0.0561$ \citep{owen97} and hosts the most powerful radio-emitting AGN at $z<0.1$ \citep[e.g.][]{Carilli96}.  Cygnus A shows a radio jet to the northwest  and a counter jet to the southeast of the nucleus, seen from sub-pc scales to up to 70 kpc from the nucleus, with superb radio lobes and strong hot-spots \citep[Figure \ref{large},][]{Perley84,Linfield85,Carilli96}. The jet and hot-spots are also seen in X-rays, together with  a giant cavity surrounding the galaxy, carved by the jet 
\citep[Figure \ref{large},][]{wilson00,wilson06,snios18} produced by an obscured AGN as indicated by the detection of absorbed, power-law X-ray emission \citep[e.g.][]{ueno94,young02}.

In the optical, Cygnus A presents a complex dust bipolar structure and a kpc-scale ionisation bicone along the southeast-northwest direction, well aligned with the radio jet  \citep[Figure \ref{large},][]{Carilli89,Jackson94,jackson96,jackson98}. The nucleus is highly obscured in the optical due to dust along the line of sight \citep[e.g.][]{Vestergaard93}, with visual extinction  of $A_{\rm v}=40-150$ mag as derived using mid \citep{Imanishi00,Ramirez14} and near-IR \citep{Djorgovski91,Packham98,Tadhunter99} observations.  \citet{Privon12} modeled the spectral energy distribution (SED) of Cygnus A from radio wavelengths to the mid-infrared, and found that the mid-infrared emission is consistent with radiation emitted by a dusty torus with size of $\sim$130 pc heated by an AGN with bolometric luminosity of $10^{12}$ L$_\odot$.  
 Recently, using Jansky Very Large Array images at 18-48 GHz, \citet{Carilli19} reported the detection of an elongated $\sim$500 pc structure, perpendicular to the radio jets and centreed on the core, providing a direct evidence o dusty torus in Cygnus A.  

\citet{Tadhunter03} studied the gas kinematics in the narrow-line region of Cygnus A using optical and near-IR long-slit spectra, obtained with the STIS instrument on the Hubble Space Telescope and the NIRSPEC instrument coupled to the Keck II Telescope, respectively. They found that the kinematics of Pa$\alpha$, H$_2$2.1218\,$\mu$m and [O\,{\sc iii}]$\lambda$5007 show a rotation pattern across the nucleus along a position angle PA$=180^\circ$, perpendicularly to the radio jet, while no evidence for rotation along the radio axis is observed. In addition, their data reveal an ionised gas outflow, northwest of the nucleus. \citet{Edwards09} used optical integral field unit observations of the inner 7.7$\times$10.8 arcsec$^2$ of Cygnus A to map the ionised gas emission structure and kinematics at an angular resolution of 0.7 arcsec. They found that the optical-emission line fluxes (H$\alpha$, [N\,{\sc ii}]$\lambda$6583 and [O\,{\sc iii}] $\lambda$6583) are consistent with emission of gas photoionised by and AGN in most locations, the flux distributions are more  elongated along the northwest-southeast direction, with the highest flux values observed to the northwest and in the central core. The gas velocity fields show  gradients of up to $\pm$200 km\,s$^{-1}$ along the northeast-southwest direction, consistent with rotation, while the line widths are large ($\sigma\approx300$ km\,s$^{-1}$) in regions co-spatial with the central radio core and to the west of the nucleus, suggesting an interaction between the radio jet and the interstellar medium \citep{Edwards09}.

The near-IR nuclear spectrum of Cygnus A presents emission lines produced in a wide range of gas phases, including lines from molecular, low and high-ionisation gas \citep{Wilman00,mcgregor07}. Gemini Near-Infrared Integral Field Spectrograph (NIFS) observations show that the emission structure of the ionised gas (traced by [Si\,{\sc x}]1.4305\,$\mu$m,  Pa$\alpha$ and [Si\,{\sc vi}]1.9630\,$\mu$m emission lines) presents a bi-conical morphology with axis along the radio jet, similar to that seen in the  broad-band K continuum \cite{Canalizo03}; the [Fe\,{\sc ii}]1.6440\,$\mu$m emission is stronger at the nucleus and presents a more rounded flux distribution, while the H$_2$2.1218\,$\mu$m emission shows extended structure approximately along the north-south direction, perpendicularly to the  radio jet axis \citep{mcgregor07}.

\begin{figure}
\includegraphics[width=0.24\textwidth]{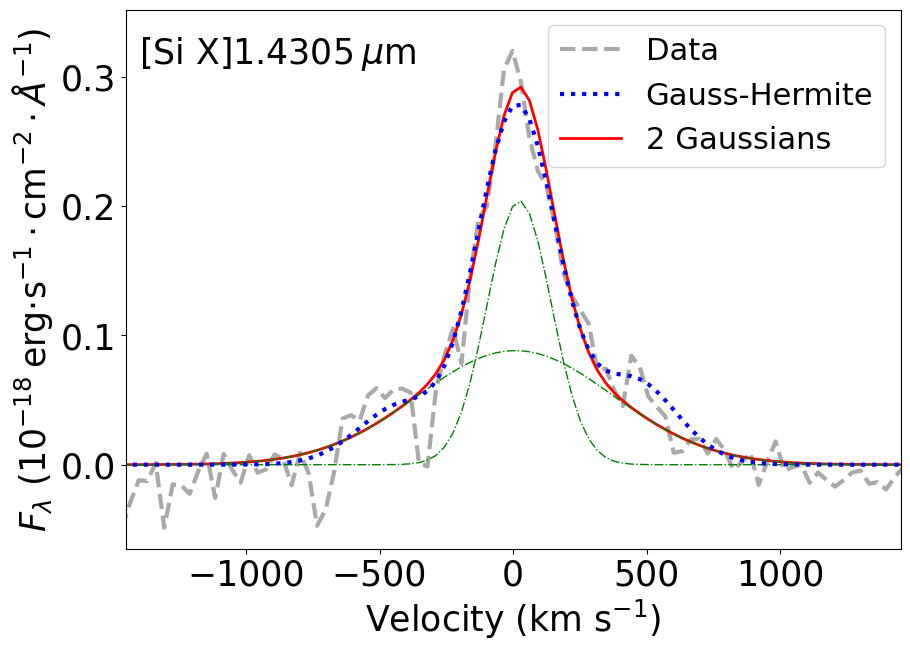}
\includegraphics[width=0.24\textwidth]{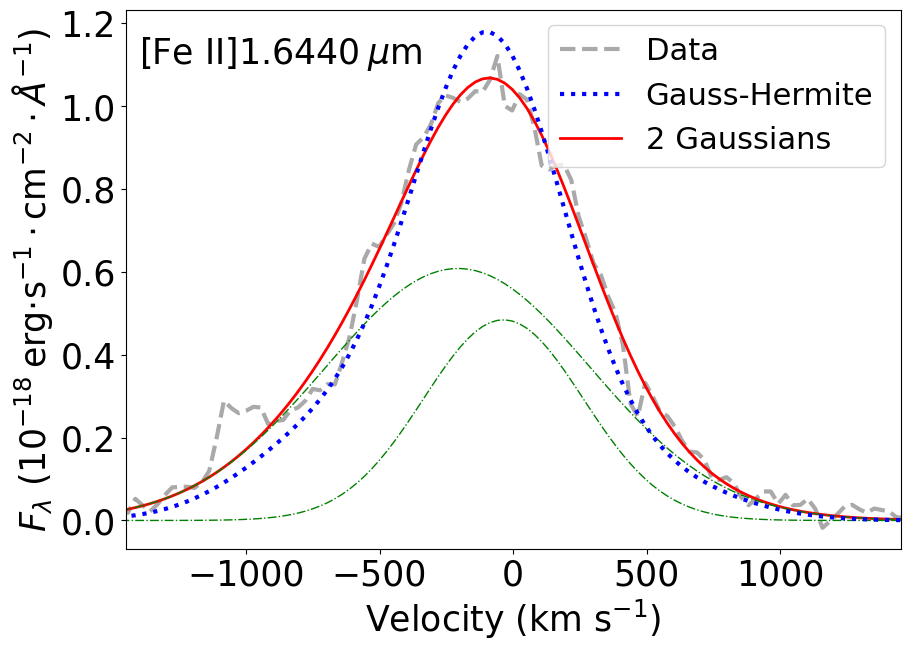}
\includegraphics[width=0.24\textwidth]{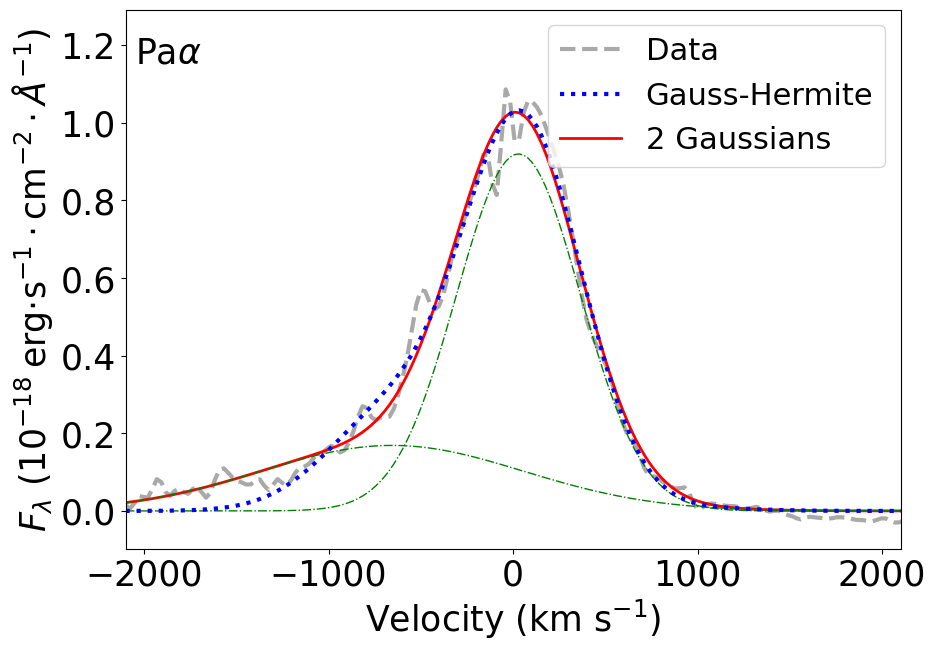}
\includegraphics[width=0.24\textwidth]{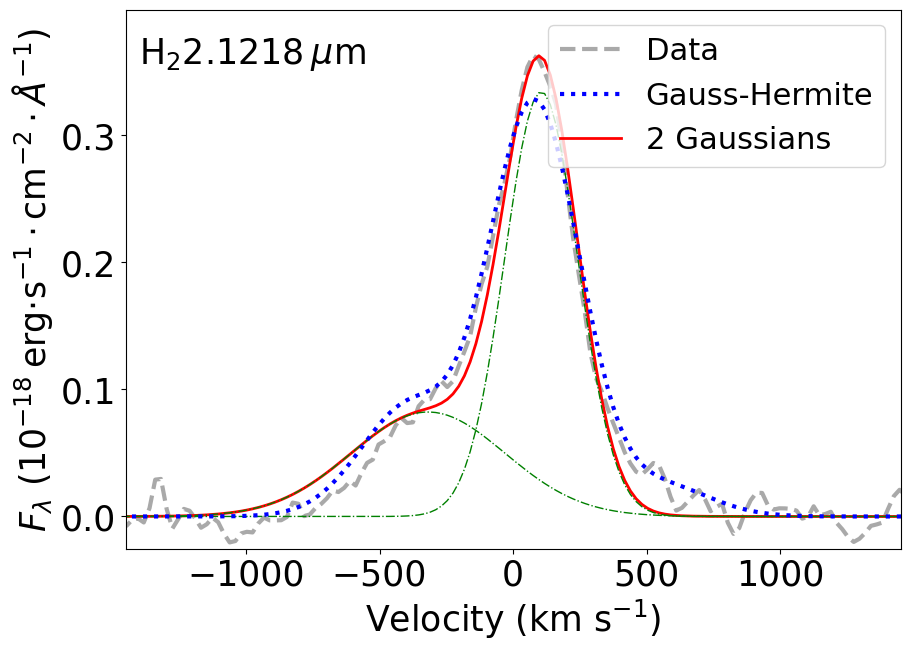}
\caption{Examples of fits of the [Si X]1.4305$\:\mu$m (top left), [Fe II]1.6440$\:\mu$m (top right), Pa$\alpha$ (bottom left)  and  H$_2$2.1218$\:\mu$m (bottom right) emission line profiles for the  nuclear spaxel. The data are shown as  gray dashed lines, the fits by Gauss-Hermite series as dotted blue lines, the two-Gaussian models as red continuous line and the individual Gaussians are shown as dashed-dotted green lines. }
\label{fig:fits}
\end{figure}

In this paper,  we map the near-IR emission line flux distributions and kinematics in the inner 3.5$\times$3.5 kpc$^2$ of Cygnus A using integral field spectroscopy, obtained with the Gemini NIFS instrument. This paper is organized as follows: Sec. \ref{datasec} describes the observations, data reduction and  measurement proceadures, our results are presented in Sec. \ref{resulsec} and discussed in Sec.~\ref{disc}. The conclusions of this work are summarized in Sec.~\ref{concsec}. Throughout this paper, we adopt a $h=0.7, \Omega_m=0.3, \Omega_{\Lambda}=0.7$ cosmology and a use the redshift of $z=0.0561$ for Cygnus A \citep{owen97}.

\section{Data and Measurements}
\label{datasec}

We analyse archival H,  K$_{\rm short}$  and K-band data obtained with the Gemini Near-Infrared Integral Field Spectrograph (NIFS), under the program GN-2006A-C-11 (PI: McGregor). NIFS has a square field of view of 3$\times$3\,arcsec$^2$, divided into 29 slices with an angular sampling of 0$\farcs$103$\times$0$\farcs$042 and is  optimized to operate with the ALTtitude conjugate Adaptive optics for the InfraRed (ALTAIR) instrument \citep{mcgregor03}. The H band observations were done using the H$_-$G5604 grating and the	JH$_-$G0602 filter, for the  K$_{\rm short}$  observations the  K$_{\rm short-}$G5606 grating and the HK$_-$G0603 filter were used and K-band data were obtained using the K$_-$G5605 grating	and the HK$_-$G0603 filter. The resulting spectral ranges are 1.476--1.802\,$\mu$m, 1.906--2.341\,$\mu$m and 2.009--2.442\,$\mu$m for the  H,  K$_{\rm short}$  and K-band, respectively. The on-source exposure time was 1.5 hours in the H band and 1.75 hours in the K$_{\rm short}$  and K bands, divided into individual exposures of 900\,sec. Preliminary emission-line flux maps were already presented in \citet{mcgregor07}, based on these data. The data reduction was accomplished using the {\sc nifs.gemini.iraf}
package, following the standard procedure as described in previous works \citep[e.g][]{rogemarN4051,rogemar_stellar}.


The angular resolution is 0.18$\pm$0.03 arcsec (corresponding to  210$\pm$35 pc at the galaxy) for the  K$_{\rm short}$ and K bands, as measured from the full-width at half maximum (FWHM) of the flux distribution of the standard star. For the H band, the angular resolution 0.16$\pm$0.03 arcsec (185$\pm$35 pc).  For both bands, the resulting velocity resolution is 48$\pm$5\,km\,s$^{-1}$,  estimated from the FWHM of emission lines of the wavelength calibration lamp spectra.

The bottom panel of Fig.\,\ref{large} shows the NIFS H+K band nuclear spectrum of Cygnus\,A, extracted within the circular aperture with 0\farcs25 radius.  We follow a similar procedure as described in \citet{rogemar_N1275} and use the {\sc ifscube} code \citep{ifscube} code to fit the emission-line profiles of [Si\,{\sc x}]1.4305\,$\mu$m, [Fe\,{\sc ii}]1.5331\,$\mu$m, [Fe\,{\sc ii}]1.6440\,$\mu$m, Pa$\alpha$, [S\,{\sc xi}]1.9201\,$\mu$m, 
Br$\delta$, H$_2$1.9576\,$\mu$m, [Si\,{\sc vi}]1.9630\,$\mu$m, H$_2$2.0338\,$\mu$m, H$_2$2.1218\,$\mu$m, Br$\gamma$, H$_2$2.2235\,$\mu$m and H$_2$2.2477\,$\mu$m, 
and measure their physical properties. 
The contribution of the underlying continuum is fitted by a third order polynomial.  

In some locations, the line profiles are complex, presenting more than one kinematic component, and are not well reproduced by a single Gaussian. Thus, we fit each  emission-line profile by (i) a single-Gaussian curve, (ii) two-Gaussian curves and by (iii) Gauss-Hermite series.  Both, the Gauss-Hermite and two-Gaussian are able to reproduce the observed profiles over the whole field-of-view. The H$_2$, [Fe\,{\sc ii}] and  H\,{\sc i} emission lines trace distinct gas phases. The H$_2$ traces the hot molecular gas ($T\approx2\,000 $K), the [Fe\,{\sc ii}] emission arises from partially ionized zones, while the H\,{\sc i}  recombination lines are produced in fully ionized gas. Considering that,  during the fit, the velocity and velocity dispersion  of the components of emission lines from the [Fe\,{\sc ii}], H\,{\sc i} and H$_2$ were tied, separately.

Figure~\ref{fig:fits} shows examples of the fits of the [Si\,{\sc x}]1.4305\,$\mu$m (top-left),  [Fe\,{\sc ii}]1.6440\,$\mu$m (top-right), Pa$\alpha$ (bottom-left),  and  H$_2$2.1218\,$\mu$m (bottom-right) line profiles for the nuclear spaxel, corresponding to the location where the lines are the most complex.  The line profiles clearly present two kinematic components: a narrow and a broad-blueshifted component. In each panel, the observed profile is shown as a gray dashed line, the fit by a Gauss-Hermite series as a blue dotted line and the two-Gaussian model as a red continuous line.  The fluxes of the emission lines obtained by modeling the profiles by Gauss-Hermite series and by Gaussian curves are consistent with each other, with a standard deviation of the difference between these fluxes smaller than 10 per cent for all considered lines.

\section{Results}
\label{resulsec}

The near-IR spectra of Cygnus A show strong emission lines, which can be used to map the multiphase gas distribution and kinematics, of the hot molecular gas (traced by the H$_2$ lines), low-ionisation gas (traced by the [Fe\,{\sc ii}] lines and H\,{\sc i} recombination lines) and coronal gas (traced by high-ionisation emission-lines species, such as [Si\,{\sc x}]1.4305\,$\mu$m and [Si\,{\sc vi}]1.9630\,$\mu$m). 

\subsection{Results from the Gauss-Hermite fits}

\begin{figure*}
\includegraphics[width=1\textwidth]{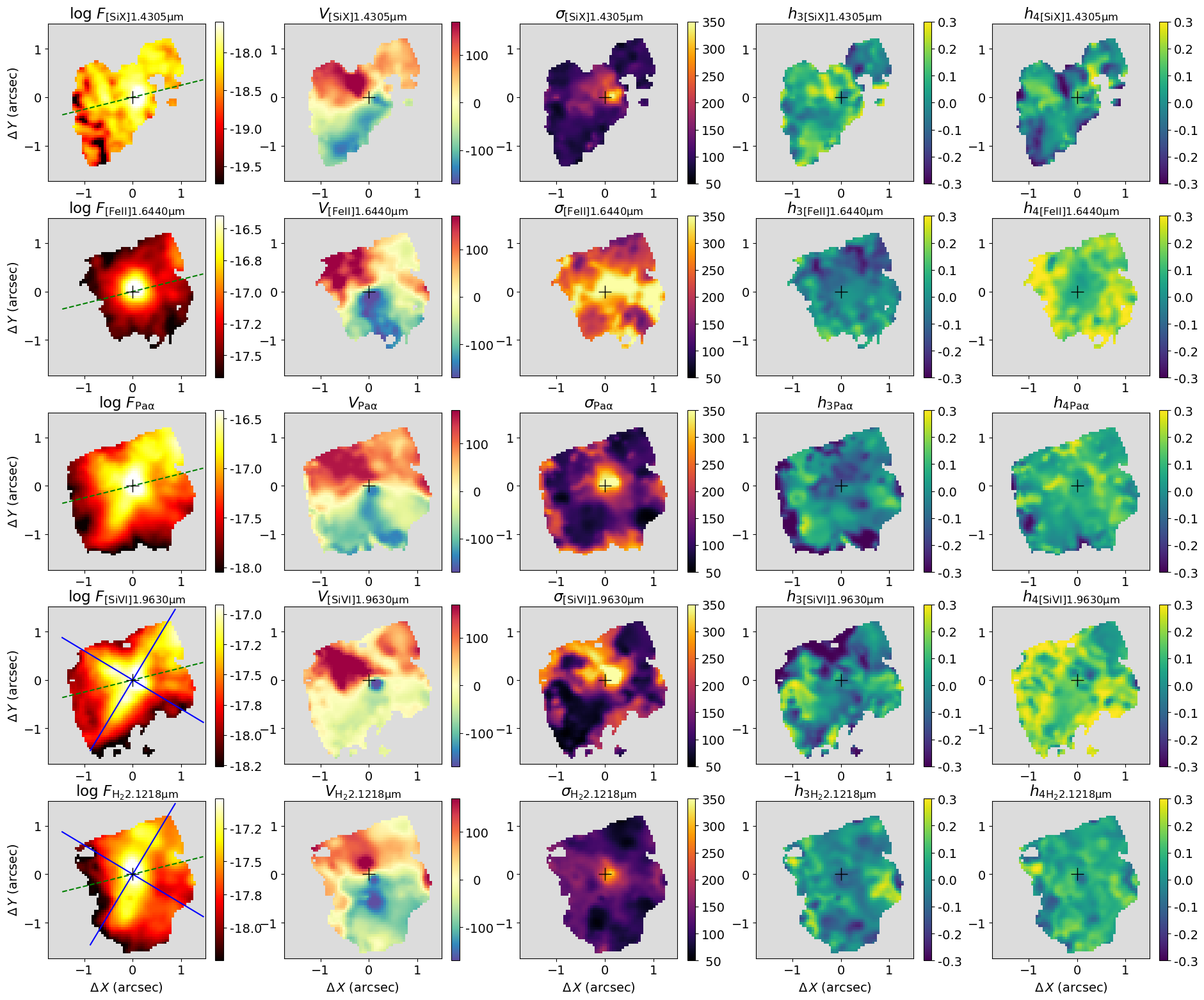}
\caption{Maps produced by modeling the emission-line profiles by Gauss-Hermite series. Each row shows the flux, velocity, $\sigma$, $h_3$ and $h_4$ maps for the emission line identified at the title of each panel.  The gray regions correspond to locations where the corresponding emission line is not detected above 3$\sigma$ level of the continuum noise.  The central cross marks the position of the peak of the K-band continuum. The green dashed line overplotted in the flux maps shows the orientation of the radio jet  \citep{Carilli96} and the blue continuous lines overlaid to the [Si\,{\sc vi}] and H$_2$  flux maps represent the bi-conical structure that delineates the highest flux levels observed in emission lines from the ionised gas. The colour bars show the  line fluxes in logarithmic units of ${\rm erg s^{-1} cm^{-2}}$ of each spaxel and the velocity and $\sigma$ in km\,s$^{-1}$. In all panels north is up and east is to the left.   }
\label{fig:gh}
\end{figure*}

Figure~\ref{fig:gh} shows maps for the fluxes (first column), velocities (second column), $\sigma$ (third column), $h_3$ (forth column) and $h_4$ moments (fifth column) obtained from the fits of the observed line profiles by Gauss-Hermite series.  We present these maps for  the [Si\,{\sc x}]1.4305\,$\mu$m,  [Fe\,{\sc ii}]1.6440\,$\mu$m, Pa$\alpha$, [Si\,{\sc vi}]1.9630\,$\mu$m and H$_2$2.1218\,$\mu$m emission lines, which present the highest signal-to-noise (S/N) ratio among their species. Gray regions in these maps represent masked locations, where  the lines are not detected above 3$\sigma$ continuum level.  

The emission-line flux distributions (first column of Fig.~\ref{fig:gh}) presented here are similar to those shown in \citet{mcgregor07}, obtained by direct integration of the line profiles within a spectral window of $\pm$500 km\,s$^{-1}$,  but the maps presented by these authors are noisier than ours due to the different measurement method.  The flux maps for [Si\,{\sc x}]1.4305\,$\mu$m,  Pa$\alpha$ and [Si\,{\sc vi}]1.9630\,$\mu$m show a well defined `X-shaped' emission morphology in the highest flux levels. This structure seems to delineate the walls of a bicone oriented along the position angle (PA) of the radio jet \citep[PA=284$^\circ$, e.g. Fig~\ref{large} and ][]{Carilli96}, with an opening angle of $\approx$90$^{\circ}$. The orientation of the radio jet is represented by the green line in Fig.~\ref{fig:gh} and the blue lines delineate the bi-conical emission structure. The [Fe\,{\sc ii}]1.6440\,$\mu$m shows a more rounded and centrally peaked flux distribution.  Some collimated emission is also observed, mainly to the north-west of the nucleus. The H$_2$ emission structure is distinct from that of the ionised gas. The highest flux levels are seen approximately along the north-south direction, outside the bi-conical emission structure observed in the ionised gas.

The velocity maps (second column of Fig.~\ref{fig:gh}) of all emission lines show redshifts to the north-northeast and blueshifts to the south-southwest of the nucleus. The velocity values range from $-170$ to 170 km\,s$^{-1}$ relative to the systemic velocity of the galaxy of $V_s=16800$ km\,s$^{-1}$, obtained by fitting the  H$_2$ velocity field by a rotating disc model (see Sec.~\ref{disc:kin}). Some differences are seen between the maps for the ionised gas and the H$_2$ velocity field. For instance, the velocity fields for the ionised gas show the highest redshifts further east ($\Delta\alpha, \Delta\delta$) $\approx$ ($-$0.5 arcsec, $+$0.3 arcsec) of the region with the highest velocities in the H$_2$ map, which presents the largest velocity gradient in the direction perpendicular to the radio jet.

The $\sigma$ maps are presented in the third column of Fig.~\ref{fig:gh} and show a wide range of values, from $\sim$50 to 350\,km\,s$^{-1}$. These maps were corrected for the instrumental broadening of $\sigma_{\rm inst}=20$\,km\,s$^{-1}$. The lowest $\sigma$ values are observed at distances $r\gtrsim$0.5 arcsec from the nucleus, mainly to south-southeast and northwest of the nucleus, while the highest values are observed at the nucleus, to  northeast and southwest of it.  Overall, the H$_2$ shows the smallest mean $\sigma$ values and the highest values are observed for the [Fe\,{\sc ii}]. 

The maps for the $h_3$ Gauss-Hermite moment are shown in the forth column of Fig.~\ref{fig:gh}. This parameter measures asymmetric deviation from a Gaussian profile, such as blue ($h_3<0$) or
red wings ($h_3>0$). In most locations, the observed values of $h_3$ are in the range from $-0.1$ to $0.1$. The lowest values ($\sim-0.2$) are seen at the nucleus and to the north-northwest, regions where emission-line profiles clearly present a blue wing (e.g. Fig.~\ref{fig:fits}). The highest values ($\sim+0.2$) are observed mainly to the east of the nucleus, indicating red wings on the line profiles.

The $h_4$ moment measures symmetric deviation from a Gaussian profile,  {\it i.e.} it quantifies the peakiness of the profile, with $h_4 < 0$ for a broader and  $h_4 > 0$ for a more peaked profile than a Gaussian \citep[e.g.][]{rogemar_profit}. The lowest values in the $h_4$ maps (fifth column of Fig.~\ref{fig:gh}) for Cygnus A are observed close to the nucleus, co-spatially with the highest $\sigma$ values, indicating that the line profiles are broader than a Gaussian.

\subsection{Emission-line flux ratios}

\begin{figure*}
\includegraphics[width=0.9\textwidth]{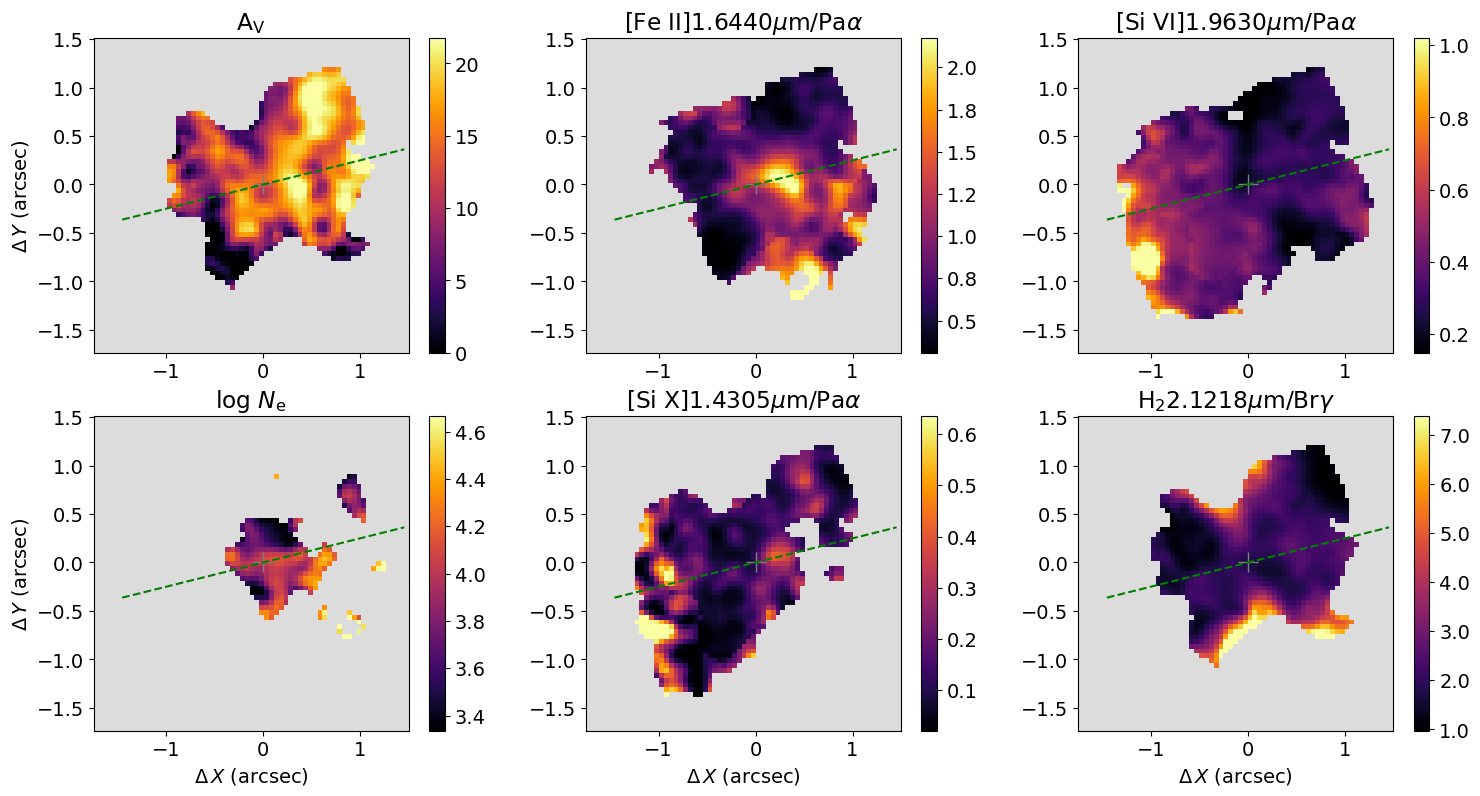}
\caption{Maps for the visual extinction, $A_{\rm v}$, electron density, $N_{\rm e}$, [Fe\,{\sc ii}]1.6440\,$\mu$m/Pa$\alpha$, [Si\,{\sc vi}]1.19630\,$\mu$m/Pa$\alpha$ [Si\,{\sc x}]1.4305\,$\mu$m/Pa$\alpha$ and H$_2$2.1218\,$\mu$m/Br$\gamma$ flux ratios. The green line shows the orientation of the radio jet \citep{Carilli96} and the gray regions correspond to masked locations where one or both lines were not detected above 3$\sigma$ level of the continuum noise.}
\label{fig:ratios}
\end{figure*}

We estimate the visual extinction ($A_{\rm v}$) using the Pa$\alpha$/Br$\gamma$ emission-line flux ratio, calculated using the fluxes measured from the modeling of the line profiles by Gauss-Hermite series. The  $A_{\rm v}$ can be estimated as follows \citep[e.g.][]{calzetti00,dominguez13,rogerio21_sfr}
\begin{equation}
    A_{\rm v}  =  \frac{2.5}{\left(f_{\lambda} ( \mathrm{Br}\gamma) - f_{\lambda} (\mathrm{Pa}\alpha)\right)}\,\log \left[\frac{(F_{\mathrm{Pa}\alpha}/F_{\mathrm{Br}\gamma})_{\rm obs}}{(I_{\mathrm{Pa}\alpha}/I_{\mathrm{Br}\gamma})_{\rm int}}\right],
\end{equation}

\noindent where $ f_{\lambda} (\mathrm{Pa}{\alpha})$ and $f_{\lambda} ( \mathrm{Br}{\gamma})$ are the reddening curve values at the $\mathrm{Pa}{\alpha}$ and $ \mathrm{Br}{\gamma}$ wavelengths, $(F_{\mathrm{Pa}\alpha}/F_{\mathrm{Br}\gamma})_{\rm obs}$ is the observed Pa$\alpha$/Br$\gamma$ emission-line flux ratio in each spaxel and $(I_{\mathrm{Pa}\alpha}/I_{\mathrm{Br}\gamma})_{\rm int}=12.4$ is the theoretical Pa$\alpha$/Br$\gamma$ intensity ratio, assuming the Case B \ion{H}{i} recombination an electron temperature of $T_e = 20\,000 $~K \citep[a typical value for the narrow line region, e.g.][]{revalski18a,revalski18b,rogemar_te21} at the low-density limit \citep{Osterbrock06}. Using the extinction law of \citet{cardelli89} we obtain $ f_{\lambda} (\mathrm{Pa}{\alpha})=0.147$ and $f_{\lambda} (\mathrm{Br}_{\gamma})=0.116$, and thus:
\begin{equation}
    A_{\rm v}  =  -80.6 \log \left[\frac{(F_{\mathrm{Pa}\alpha}/F_{\mathrm{Br}\gamma})_{\rm obs}}{12.4}\right].
\end{equation}
The $A_{\rm v}$ map for Cygnus A is shown in the left panel of Fig.~\ref{fig:ratios}. High $A_{\rm v}$ values are observed in most location, reaching values of up to 20 mag to the northwest of the nucleus. The mean value and standard deviation of the mean are 12.5 mag and 6.3 mag, respectively.

The  [Fe\,{\sc ii}]1.5330\,$\mu$m/[Fe\,{\sc ii}]1.6440\,$\mu$m emission-line ratio can be used to estimate the electron density, $N_{\mathrm {e}}$  \citep[e.g.][]{sb_n4151}. We correct the line fluxes by extinction using the $A_{\rm v}$ values measured for each spaxel and the extinction law of \citet{cardelli89} and  derive $N_{\mathrm {e}}$ using the {\sc PyNeb} routine \citep{luridiana15},  assuming an electron temperature of $T_e = 20\,000 $~K \citep[e.g.,][]{revalski18a,revalski18b,rogemar_te21}. The [Fe\,{\sc ii}]1.5330\,$\mu$m emission-line is only detected in the inner $\sim$0.5 arcsec and in some spaxels to the northwest of the nucleus, allowing the measurements of $N_{\rm e}$ only in these regions. 
The electron density map, shown in the bottom left panel of Fig.~\ref{fig:ratios}, presents values in the range $3.4 \lesssim \log N_{\rm e}/{\rm (cm^{-3})} \lesssim 4.6$, with an average value of $\langle \log N_{\rm e}/{\rm (cm^{-3})} \rangle =3.9\pm0.3$.  

The emission-line intensity ratio maps shown in Fig.~\ref{fig:ratios} were constructed after correcting the fluxes by extinction using the $A_{\rm v}$ values measured for each spaxel and the extinction law of \citet{cardelli89}. These maps are useful to investigate the gas excitation \citep[e.g.][]{reunanen02,ardila04,rogerio13,rogemar21_exc}. The excitation mechanisms of the H$_2$ and [Fe\,{\sc ii}] near-IR lines can be investigated using the [Fe\,{\sc ii}]1.6440\,$\mu$m/Pa$\alpha$ and H$_2$2.1218\,$\mu$m/Br$\gamma$ flux ratios.  The highest values of [Fe\,{\sc ii}]/Pa$\alpha$ are observed at the nucleus and to the southwest, while the smallest ratios are seen mainly at distances larger than 0.5 arcsec from the nucleus. The H$_2$/Br$\gamma$ map presents the lowest values within the bi-conical structure seen in the flux distributions of the ionised gas emission lines, while the highest values are observed approximately perpendicularly to this structure. The [Si\,{\sc x}]1.4305\,$\mu$m and [Si\,{\sc vi}]1.19630\,$\mu$m   
are coronal lines with ionisation potentials of 351.1  and 166.8 eV, respectively \citep{ardila11} and thus they are  tracers of the high ionisation gas. The [Si\,{\sc vi}]/Pa$\alpha$ ratio map presents the highest values (of up to 1.0) to the southeast and the smallest values ($\approx$0.2) are observed to the northwest of the nucleus.  A similar behaviour is observed in the [Si\,{\sc x}]1.4305\,$\mu$m/Pa$\alpha$ map, with values of up to $\sim$0.5 observed to the southeast of the nucleus and smaller values to the northwest of it.

\subsection{Results from the Gaussian fits}

\begin{figure*}
\includegraphics[width=0.9\textwidth]{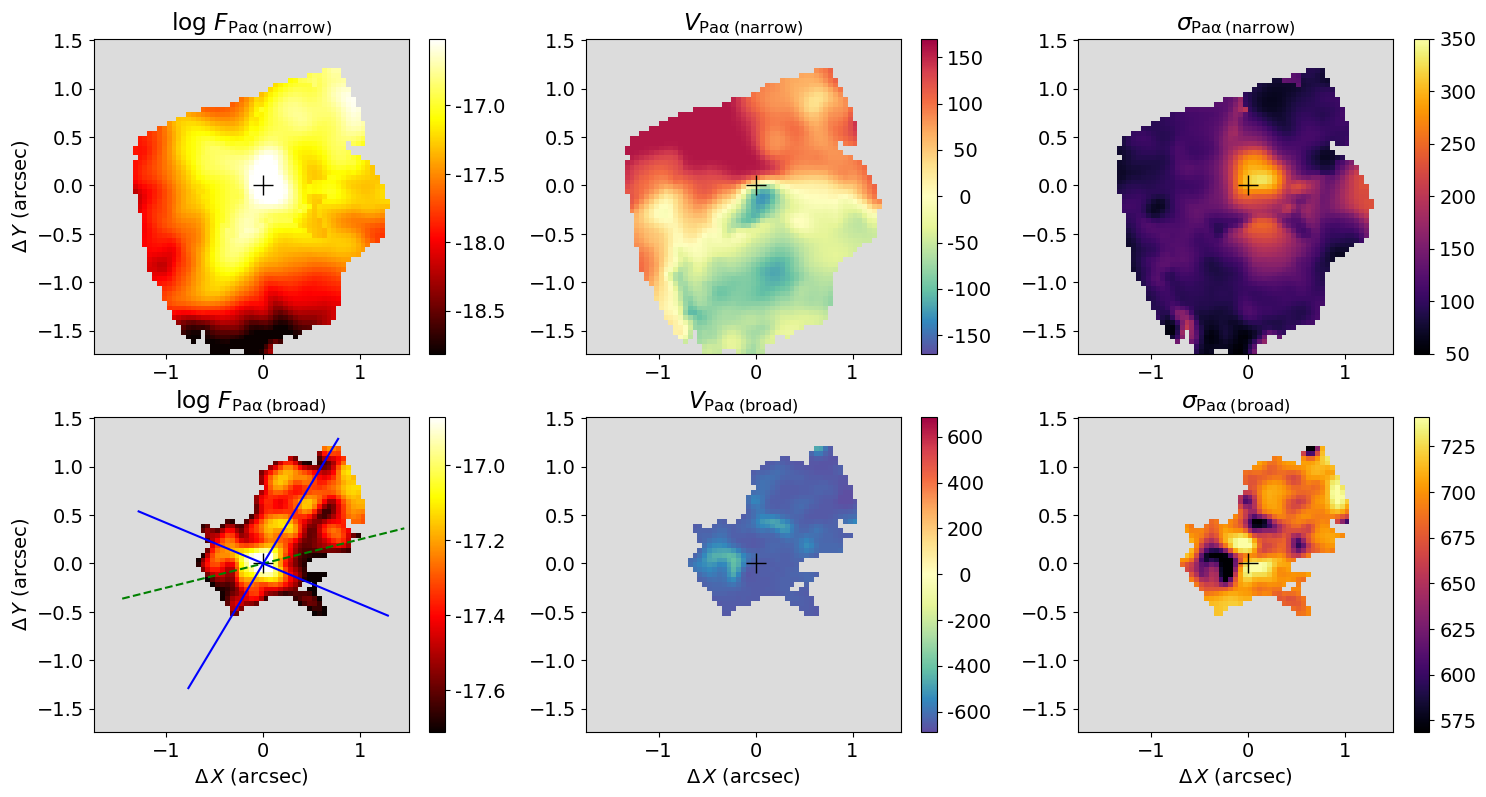}
\caption{Flux (left), centroid velocity (middle) and velocity dispersion (right) maps for the narrow (top panels) and broad (bottom panels) for Pa$\alpha$. The green dashed line and the blue continuous lines overplotted in the flux map for the broad component show the orientation of the radio jet  \citep{Carilli96} the bi-conical structure seen in the ionised gas emission, respectively. The colour bars show  the fluxes in logarithmic units of ${\rm erg s^{-1} cm^{-2}}$ of each spaxel and the velocity and $\sigma$ in km\,s$^{-1}$. In all panels north is up and east is to the left.  
}
\label{fig:2GPaa}
\end{figure*}

\begin{figure*}
\includegraphics[width=0.9\textwidth]{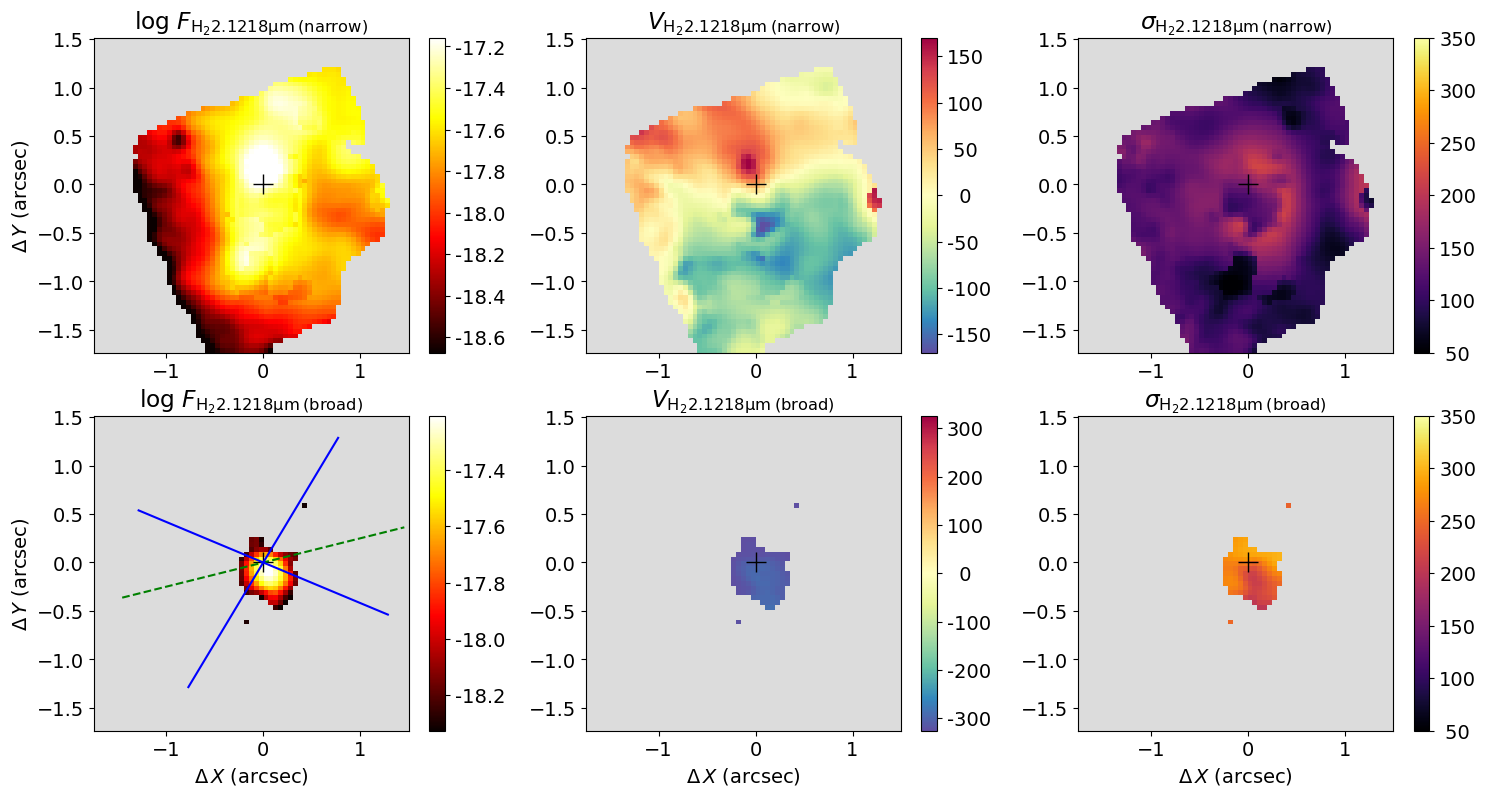}
\caption{Same as Fig.~\ref{fig:2GPaa}, but for the H$_2$2.1218\,$\mu$m emission line.}
\label{fig:2GH2}
\end{figure*}

All the emission lines present in the Cygnus A H+K spectra show a broad and blueshifted component observed in some locations, particularly strong in the central region. In Figs.~\ref{fig:2GPaa} and \ref{fig:2GH2} we show the flux (left panels), centroid velocity (middle panels) and velocity dispersion (right panels) maps for the narrow (top) and broad (bottom) components of Pa$\alpha$ and H$_2$2.1218\,$\mu$m emission lines, respectively.  These lines were chosen because they present the highest S/N ratios among the lines produced by the ionised and molecular gas, respectively. We mask out spaxels where the amplitude of the fitted components are smaller than 3$\sigma$ level of the continuum noise, next to the considered line. In locations where the broad component is not detected, we fit the corresponding line profile by a single Gaussian function, representative of the narrow component. 

The maps for the narrow component are similar to those shown in Fig.~\ref{fig:gh}, obtained by fitting the line profiles by Gauss-Hermite series.   The flux distribution for the Pa$\alpha$ broad component shows a linear structure to the northwest of the nucleus, extending to up to 1.5 arcsec (1.75 kpc) from it. The peak of emission is seen at the galaxy's nucleus. A similar extended structure is also seen in [Fe\,{\sc ii}]1.6440\,$\mu$m, while for emission lines from highly ionised gas (e.g. [Si\,{\sc x}]1.4305\,$\mu$m) only the nuclear emission is detected above the 3$\sigma$ level of the continuum noise.  The Pa$\alpha$ broad component is blueshifted, with typical velocities of 400--600\,km\,s$^{-1}$ relative to the systemic velocity of the galaxy, and presents $\sigma=600-720$\,km\,s$^{-1}$. The broad component in H$_2$2.1218\,$\mu$m is detected only at the nucleus and has a smaller centroid velocity and velocity dispersion than those of Pa$\alpha$. This component is blueshifted by $\sim$300\,km\,s$^{-1}$ and has a $\sigma\approx280$\,km\,s$^{-1}$.

\section{Discussion}\label{disc}

\subsection{Gas extinction }

A complex dust structure in the inner few kpc of Cygnus A is observed in optical broad band images \citep[e.g.][]{Carilli89,Jackson94,jackson96,jackson98} and the nuclear visual extinction has been measured using optical \citep[e.g.][]{Edwards09}, near-IR \citep{Djorgovski91,Packham98,Tadhunter99,Wilman00} and mid-IR \citep{Imanishi00,Ramirez14} spectroscopy. These studies show that the visual extinction ($A_{\rm v}$) values based on optical emission lines are smaller than those derived using infrared lines. For instance, \citet{Tadhunter94} obtained $A_{\rm v}\sim3$ mag using the H$\alpha$/H$_\beta$ flux line ratio and \citet{Wilman00} found $A_{\rm v}=9$ mag based on the Br$\gamma$/Pa$\alpha$ ratio for the nucleus of Cygnus A.  Our extinction map (Fig.~\ref{fig:ratios}) shows values of up to 20 mag, with an average value of 12.5$\pm$6.3 mag.  \citet{Edwards09} presented an extinction map of the inner $4.5\times4.5$ arcsec$^2$ region of Cygnus A, based on the H$\alpha$/H$_\beta$ flux line ratio observed using integral field spectroscopy at an angular resolution of 0.7\,arcsec. Their map, although noisy, show the highest value of  $A_{\rm v}\sim3$ mag along the northwest-southeast direction. Our map shows a similar behaviour, but with overall higher values, with the highest gas extinction seen along the walls of the ionisation cone.

As already discussed in \citet{Wilman00}, a likely explanation for the higher visual extinction derived using near-IR emission lines than those based on optical lines is that the dust is not uniformly distributed and mixed with the gas.  The extinction in the near-IR is lower than in the optical, and thus, infrared observations can penetrate more deeply into the gas emission structure and probe more obscured regions. Indeed, higher $A_{\rm v}$ values based on near-IR emission lines than those from optical emission lines are commonly reported in the literature for nearby galaxies \citep[e.g.][]{Martins13}.

\subsection{Gas Density}\label{sec:Ne}

 The [Fe\,{\sc ii}] near-IR emission lines originate in partially ionised zones,  which can be produced by X-rays \citep{simpson96} or shocks \citep{forbes93}, in the central region of galaxies. Similarly, the [S\,{\sc ii}]$\lambda\lambda$6717,6731 emission lines also originate in partially ionised zones, but the critical density for the [S\,{\sc ii}] lines ($2.5\times10^4$ cm$^{-3}$) is smaller than that for the [Fe\,{\sc ii}] lines ($\sim 10^5$ cm$^{-3}$; \citealt{sbN4151Exc}).  For Cygnus A, the average electron density in the [Fe\,{\sc ii}] emission region is $N_{\rm e}=7900\,{\rm cm^{-3}}$ (Fig.~\ref{fig:ratios}), above the critical density for the [S\,{\sc ii}].

 \citet{Tadhunter94} reported values of $1\langle\:$[S\,{\sc ii}]$\lambda6717$/[S\,{\sc ii}]$\lambda6731\:\langle1.4$  using long-slit slit observations along two position angles, PA=15$^\circ$ and PA=105$^\circ$, covering the inner 3 arcsec of Cygnus A. These values correspond to densities of $25{\, \rm cm^{-3}}\: \langle N_e\:\langle 915{\, \rm cm^{-3}}$ as obtained using the  {\sc PyNeb} routine \citep{luridiana15} and  assuming an electron temperature of $T_e = 20\,000 $~K.  Using the fluxes of the [S\,{\sc ii}] doublet obtained from the nuclear spectrum of Cygnus A presented by \citet{Torrealba12} and adopting the same procedure, the resulting density is $N_e\approx525{\, \rm cm^{-3}}$. Thus, the $N_e$ values in the [Fe\,{\sc ii}] emission region  is about 15 times larger than that obtained from [S\,{\sc ii}] lines.  
 
 The densities derived from the [S\,{\sc ii}] doublet seem to underestimate the density of ionised outflows in the narrow-line region of luminous Seyfert galaxies by up to two orders of magnitude, because the assumption that the  [S\,{\sc ii}]-based electron density traces the hydrogen density is invalid, as the [S\,{\sc ii}] emission arises from partially ionised zones \citep{Baron19,davies20}. An alternative to derive the density of the ionised gas is by using the optical [\ion{Ar}{iv}]$\lambda\lambda$4711,4740 emission lines, which trace denser gas phases than the [\ion{S}{ii}] line ratio. Measuring the fluxes of these lines from the nuclear spectrum of \citet{Torrealba12}, using the {\sc PyNeb} routine and adopting $T_e = 20\,000 $~K, we derive $N_e\approx4450{\, \rm cm^{-3}}$, which is roughly half the value obtained from the [Fe\,{\sc ii}] lines.   
 A possible explanation for the large discrepancy between the $N_e$ values derived from optical emission lines and from the near-IR [Fe\,{\sc ii}] is that, as mentioned above, the near-IR observations are able to probe dustier gas clouds. These clouds are likely more directly illuminated by the AGN radiation field, and possibly compressed by AGN winds, increasing the gas density.

\subsection{Flux distributions}
Ground based and Hubble Space Telescope narrow-band images of Cygnus A revealed a well defined bipolar emission structure in the inner $\sim$3 kpc \citep[e.g.][]{Pierce86,jackson96,jackson98}, which can be described as having a biconical or parabolic/hourglass morphology. This emission structure is well aligned with the orientation of the radio jet, with a position angle  of 284$^\circ$ on the sky \citep[e.g.][]{Perley84,Carilli96}; it may result from a combination of mechanical excitation of the gas in shocked regions and photoionisation by a hidden quasar like nucleus \citep[e.g.][]{jackson96}. Similar hourglass morphologies has been observed for the narrow-line region of closer Seyfert galaxies, as for instance in NGC\,4151 \citep[e.g.][]{sb_n4151} and NGC\,1068 \citep[e.g.][]{barbosa14}.

The broad-band K continuum image of the central region of Cygnus A, obtained with the Keck II Telescope at an angular resolution of 0.05 arcsec,  also clearly show the ionisation/scattering bicone and reveals an unresolved nucleus at the cones apex \citep{Canalizo03}. The NIFS K-band continuum image, shown in Fig.~\ref{large}, presents a similar morphology.  

Preliminary flux maps from the NIFS data analysed here, obtained by integrating the line profiles over $\pm$500 km\,s$^{-1}$ equivalent
waveband, were already presented in \citet{mcgregor07}. They presented flux maps for the H$_2 2.1218\,\mu$m, [Fe\,{\sc ii}]$1.6440\,\mu$m, Pa$\alpha$,  [Si\,{\sc x}]1.4305\,$\mu$m and [Si\,{\sc vi}]1.9630\,$\mu$m. Our flux maps (left panels of Fig.~\ref{fig:gh}) are well consistent with theirs, but less noisy, as their maps were constructed by direct integration of the line profiles and thus more sensitive to the continuum noise. The biconical morphology is clearly seen in the flux maps for the emission lines from the ionised gas. The gas emission seems to be tracing the walls of the cones, as seen in the K-band continuum image, delineating an "X-shaped'' emission structure. The [Si\,{\sc vi}]/Pa$\alpha$ and [Si\,{\sc x}]/Pa$\alpha$ intensity ratio maps show the highest values in the easterly cone, indicating that the ionisation parameter of the gas in this side is higher than that in the westerly cone.  A possible interpretation for this behaviour is given by \citet{mcgregor07}, in which the easterly cone is matter bounded and in the westerly cone dense matter seems to be obtruding into the cone.  

The [Fe\,{\sc ii]}$1.6440\,\mu$m shows a more round flux distribution than those of lines from other ionised species, and the highest [Fe\,{\sc ii}]/Pa$\alpha$ values are observed at the nucleus and along the northeast-southwest direction. These higher [Fe\,{\sc ii}]/Pa$\alpha$ values outside the ionisation cone may be tracing emission from shock-ionised gas produced by nuclear outflows or by the interaction of the radio jet with the gas of the disc \citep[e.g.][]{rogemar_n5929_let,rogemar_te21,lena15,couto16,Venturi21}, as shocks can be easily observed outside the ionisation cones, where  the AGN radiation field is shielded by the nuclear dusty torus \citep{zakamska14,rogemar_te21}.  

The H$_2$ flux distribution (bottom-left panel of Fig.~\ref{fig:gh}) is distinct from those of the ionised gas. Most of the H$_2$ emission is seen from locations outside the ionisation cone, extending along the north-south direction. The H$_2$ seems to be tracing the emission of gas along the major axis of the rotation disc, previously reported using near-IR \citep{Tadhunter03} and optical observations \citep{Edwards09}. 

\subsection{The [Fe\,{\sc ii}] and H$_2$ emission}

\begin{figure}
\includegraphics[width=0.9\columnwidth]{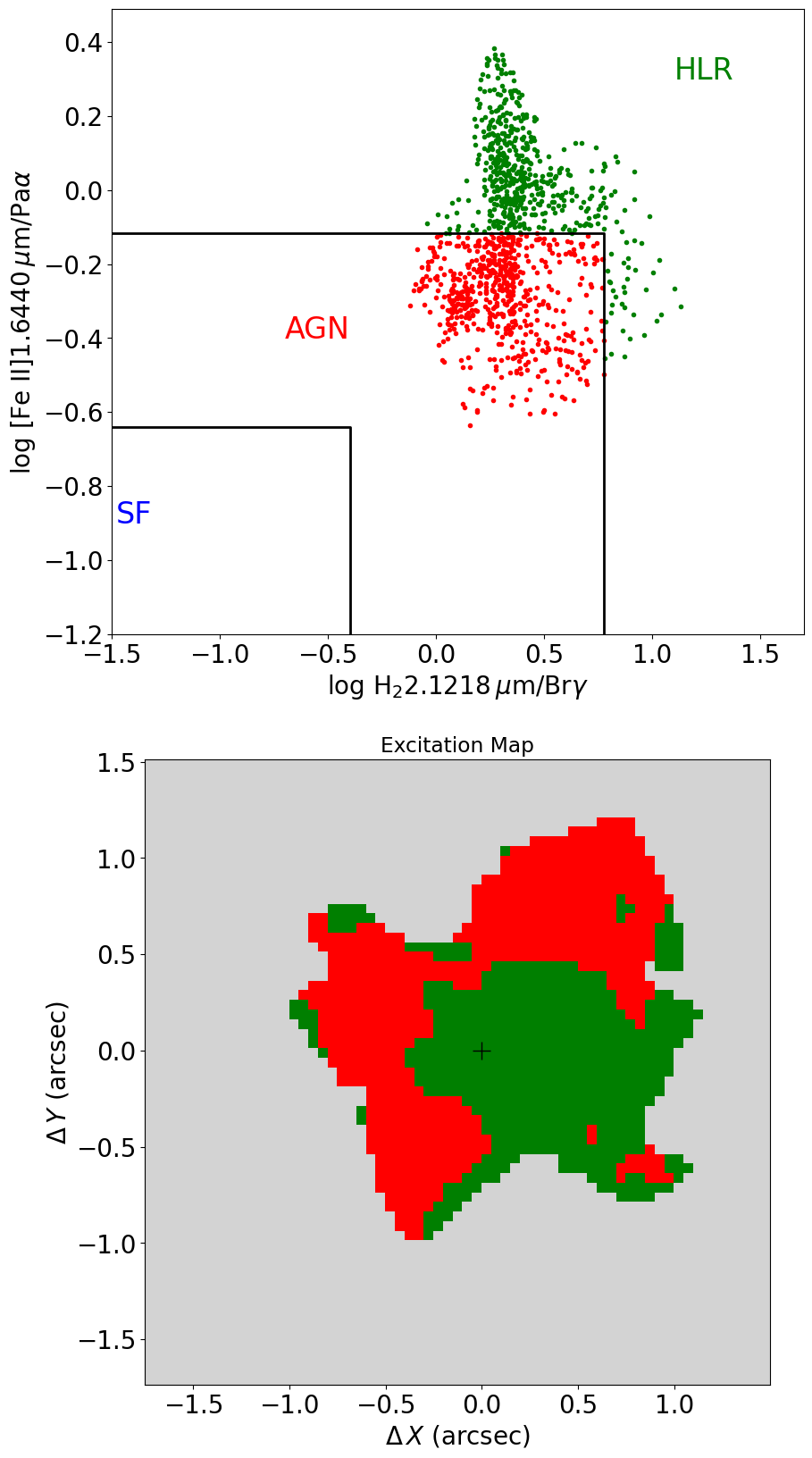}
\caption{Top: [Fe\,{\sc ii}]$1.6440\,\mu$m/Pa$\alpha$ vs. H$_2 2.1218\,\mu$m/Br$\gamma$ diagnostic diagram for Cygnus A. The lines delineating the star-forming (SF), AGN and high line ratio (HLR) regions are from \citet{rogerio13}. 
Bottom: excitation map colour-coded according to the region diagnostic diagram. }
\label{fig:FeIIdiag}
\end{figure}

Figure~\ref{fig:FeIIdiag} shows the [Fe\,{\sc ii}]$1.6440\,\mu$m/Pa$\alpha$ vs. H$_2 2.1218\,\mu$m/Br$\gamma$ diagnostic diagram for Cygnus A and the corresponding excitation map, which are useful to investigate the origin of the [Fe\,{\sc ii}] and H$_2$  emission lines \citep[e.g.][]{ardila05,rogerio13,colina15,rogemar21_exc}.  
The lines delineating the star-forming (SF), AGN and high line ratio (HLR) regions are from \citet{rogerio13}, by converting the [Fe\,{\sc ii}]$1.2570\,\mu$m/Pa$\beta$ to [Fe\,{\sc ii}]$1.6440\,\mu$m/Pa$\alpha$ using Pa$\alpha$/Pa$\beta$=2.0, assuming the Case B \ion{H}{i} recombination, an electron temperature of $T_e = 20\,000 $~K, and [Fe\,{\sc ii}]$1.6440\,\mu$m/[Fe\,{\sc ii}]$1.2570\,\mu$m=0.765 \citep{Colina93}. \citet{colina15} used VLT SINFONI observations of 10 luminous infrared galaxies to investigate the two-dimensional ionisation structure and found that in most galaxies, the [Fe\,{\sc ii}] and H$_2$ emission are correlated, while \citet{rogemar21_exc} found correlations between the  [Fe\,{\sc ii}]$1.2570\,\mu$m/Pa$\beta$ and H$_2 2.1218\,\mu$m/Br$\gamma$ line ratios in 3 of 5 Seyfert galaxies with extended emission, using NIFS observations.  In addition, \citet{rogemar21_exc} found that the HLR region in Seyfert galaxies is likely produced by shock excitation, as indicated by a correlation between the line ratios and widths. For Cygnus A, we find that all spaxels are in the AGN and HLR regions of the [Fe\,{\sc ii}]$1.6440\,\mu$m/Pa$\alpha$ vs. H$_2 2.1218\,\mu$m/Br$\gamma$ diagnostic diagram, with the HLR region extending from the nucleus to southwest and the typical AGN ratios observed along the ionisation cone.  The HLR region is mostly due to high [Fe\,{\sc ii}]/Pa$\alpha$, since the vast majority of H$_2$/Br$\gamma$ are lower than the upper limit for the AGN region of the diagram,  but this line ratio can also be increased by shocks \citep[e.g.][]{sb_n4151,rogemar21_exc}.  Thus, the [Fe\,{\sc ii}] emission in the HLR region is likely produced by shocks, as already discussed above. Alternatively, the HLR region could be produced by an enhancement of the Fe/O abundance as suggested by photoionisation models \citep[e.g.][]{dors12}.

\begin{figure}
\includegraphics[width=0.9\columnwidth]{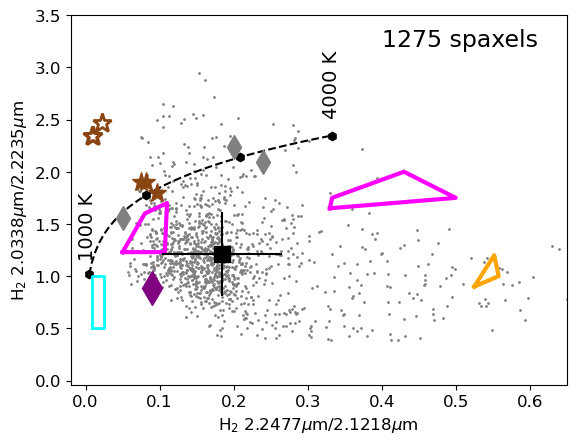}
\caption{H$_2$\,2--1\,S(1)$2.2477\,\mu$m/1--0\,S(1)$2.1218\,\mu$m vs. 1--0\,S(2)$2.0338\,\mu$m/1--0\,S(0)$2.2235\,\mu$m diagnostic diagram for Cygnus A. The small points show the ratios in individual spaxels, where the lines are detected above 3$\sigma$ of the continuum noise. The black square with error bars show the mean line ratios with their standard deviations. The other symbols are model predictions (see text). }
\label{fig:H2diag}
\end{figure}

In Fig.~\ref{fig:H2diag}, we present the H$_2$\,2--1\,S(1)$2.2477\,\mu$m/1--0\,S(1)$2.1218\,\mu$m vs. 1--0\,S(2)$2.0338\,\mu$m/1--0\,S(0)$2.2235\,\mu$m diagnostic diagram for Cygnus A, which can be used to investigate the excitation mechanisms of the H$_2$ emission lines. The small points show the ratios for each spaxel and the black square shows the mean ratios, with the error bars corresponding to the standard deviations of each ratio. The observed ratios can be compared to different model predictions. The black dashed curve represents the predicted ratios for an isothermal and uniform density gas distribution with temperatures in the range 1000 -- 4000\,K.  The purple and gray diamonds represent the predictions of shock models of \citet{Kwan77} and \citet{Smith95}, respectively. 
The filled and open brown  stars are the predictions from the thermal X-ray models of \citet{draine90} and \citet{Lepp83}, respectively. The magenta polygons show the regions covered by the AGN photoionisation models of \citet{dors12}. The range of ratios predicted by the non-thermal UV excitation models of \citet{black87} is shown as an orange polygon. The cyan rectangle delineates the predictions from the thermal UV excitation models of \citet{Sternberg89}. 

As seen in Fig.~\ref{fig:H2diag}, for Cygnus A,  most of the points in the H$_2$\,2--1\,S(1)$2.2477\,\mu$m/1--0\,S(1)$2.1218\,\mu$m vs. 1--0\,S(2)$2.0338\,\mu$m/1--0\,S(0)$2.2235\,\mu$m  diagram lie close to the shock and X-ray excitation models predictions. 
This result, along with the fact that there are no points in the SF region of Figure~\ref{fig:FeIIdiag}, indicates that  H$_2$ near-IR emission in the central region of Cygnus A is produced by thermal processes. The most likely H$_2$ excitation mechanism in Cygnus A are X-rays from the central AGN, as already discussed by \citet{Wilman00} using long slit spectroscopy and comparing the observed fluxes with the predictions from models of X-ray dissociation regions of \citet{maloney96}.  Some contribution from shocks to the H$_2$ emission can not be discarded, especially to the southwest, where the highest [Fe\,{\sc ii}]$1.6440\,\mu$m/Pa$\alpha$ and H$_2 2.1218\,\mu$m/Br$\gamma$ are observed \citep{rogemar21_exc}.

\subsubsection{The coronal line region}

As observed in Fig.~\ref{large}, the H and K band spectra of Cygnus A present strong emission lines from highly ionisation species, including  [Si\,{\sc x}]1.4305\,$\mu$m,  [S\,{\sc xi}]1.9201\,$\mu$m, 
and  [Si\,{\sc vi}]1.9630\,$\mu$m. The ionisation potentials of the parent ions of these lines are 351.1, 447.1 and 166.8 for Si\,{\sc x},  S\,{\sc xi} and  Si\,{\sc vi}, respectively \citep{ardila11}. These lines are called coronal lines (CLs), defined as those produced by ions with ionisation potentials above 100\,eV.

The origin of the CL emission in AGN is still under debate.  The nuclear ($\sim$10s pc) CL emission in AGN seems to originate in photoionised gas in the inner edge of the dusty torus \citep{shields75,korista89,ferguson97,glidden16,ardila11}, while photoionisation models fail to reproduce the CL intensities on scales of a few hundred parsec, and shocks play an important role in the production of these lines \citep{osterbrock65,ardila20}. A combination of both processes, shocks and photoionisation, can also occur  \citep[e.g.][]{viegas89,contini01,dors21}.

Near-IR, adaptive optics observations of nearby active galaxies can be used to map the extension of the CL region.  \citet{prieto05} found extensions of the [Si\,{\sc vii}]2.48\,$\mu$m from 30 to 200\,pc using adaptive optics narrow band images of four nearby Seyfert galaxies, obtained with the ESO/VLT. \citet{ms11} found sizes from 8 to 150\,pc measured from the [Si {\sc vi}]1.9650\,$\mu$m flux distributions in a sample of seven luminous Seyfert galaxies. \citet{rogemar21_exc} used Gemini NIFS data of six nearby  luminous Seyfert galaxies and found [S\,{\sc ix}]1.2523\,$\mu$m extended emission up to 80--185 pc from the nucleus. Recently, \citet{ardila20} reported highly ionised gas  emission  (traced by the [Fe\,{\sc vii}]6087\,$\mu$m) to up to 700 pc from the nucleus of the Circinus galaxy, using MUSE/VLT observations.

In Cygnus A, coronal line emission is observed up to the borders of the NIFS field of view ($\sim$1.5 arcsec/1.75 kpc) and well aligned with the radio jet,  as seen in Fig.~\ref{fig:gh}.  To the best of our knowledge, this is the most extended coronal line region reported for active galaxies, using near-IR observations.  Similarly to the result found for Circinus by \citet{ardila20}, where the extended coronal emission is likely the remnant of shells inflated by the passage of a radio jet, the extended CL emission in Cygnus A may be originated by shocks produced by the radio jet or nuclear outflows, as a pure photoionisation scenario fails at reproducing the CLs intensities at these scales \citep[e.g.][]{ardila06,mazzalay13,ardila20}.

\subsection{Kinematics}\label{disc:kin}

Previous optical and near-IR spectroscopic observations of the central region of Cygnus A reveal multiple kinematic components, including gas rotation in 100s pc and kpc scales \citep[e.g.][]{Tadhunter03,Edwards09}, ionised gas outflows \citep[e.g.][]{Tadhunter94,Wilman00,taylor03} and inflows of neutral \citep{conway95} and molecular \citep{bellamy04} hydrogen. 

Regarding the near-IR emission lines, for instance, \citet{bellamy04} presented long-slit spectroscopic data along the orientation of the radio jet (PA=105$^\circ$), obtained with the  NIRSPEC spectrograph on the Keck II telescope. They found that the Pa$\alpha$ line in the central aperture (540 pc width) presents a narrow ($\sigma\sim$100 km\,s$^{-1}$) component and a broad ($\sigma\sim$320 km\,s$^{-1}$) component, while at  1.880 kpc northwest of the nucleus, two components narrow components with $\sigma\sim$150 km\,s$^{-1}$ are observed, one blueshifted by $\sim$240 km\,s$^{-1}$ and the other redshifted by $\sim$150 km\,s$^{-1}$, relative to the rest frame of the galaxy. Our NIFS data reveal a broader blueshifted component, seen not only in the nucleus, but extending to northwest of the it (Fig.~\ref{fig:2GPaa}). A similar component is seen in [Fe\,{\sc ii}]1.6440\,$\mu$m, while for emission lines from highly ionised gas (e.g. [Si\,{\sc x}]1.4305\,$\mu$m) the broad component is detected only in the nucleus.  \citet{Wilman00} also reported a broad component in [Fe\,{\sc ii}]1.6440\,$\mu$m in the nuclear region of Cygnus A, attributed to shocks produced by the radio jet.  For the H$_2$ lines, \citet{bellamy04} detected three kinematic components in the nuclear aperture and two components at distances of 0.8 to 2 kpc from the nucleus to the northwest, along PA=105$^\circ$. They interpreted the redshifted ($\sim$225 km\,s$^{-1}$) component as being due to a molecular cloud falling through the nucleus of Cygnus A. The NIFS FoV does not cover the extra-nuclear region discussed by \citet{bellamy04}, while at the nucleus the H$_2$2.1212\,$\mu$m emission line is reproduced by two Gaussian components (Fig.~\ref{fig:fits}).  

The Gemini NIFS data allow us to spatially resolve the gas emission structure and kinematics in the inner 3.5$\times$3.5 kpc$^2$ of Cygnus A, and thus, further investigate the origin of the molecular and ionised gas emission at the probed scales. The emission-line profiles in all spaxels are reproduced by two-Gaussian functions, a narrow component which is mainly due to emission of gas in a rotating disc and a broad and blueshifted component, produced by outflows. In what follows, we discuss these kinematic components in more details. 

\subsubsection{The rotating disc}

The velocity fields of optical emission lines, obtained from IFU observations of the the inner $\sim$4 kpc of Cygnus A, show a velocity gradient of $\pm$200\,km\,s$^{-1}$ along the northeast-southwest direction, consistent with a rotation disc component \citep{Edwards09}. Our velocity maps, obtained from the Gauss-Hermite fits (Fig.~\ref{fig:gh}) and from the narrow Gaussian component (Figs.~\ref{fig:2GPaa} and \ref{fig:2GH2}), are similar to those of the optical lines.  Deviations from pure rotation are clearly observed in the velocity fields for the ionised gas, as for instance, the redshifts to the east of the nucleus, clearly seen in the velocity fields for the coronal lines. On the other hand, the H$_2$ velocity field is the one that most resembles a pure rotation pattern, with the line of nodes oriented approximately perpendicular to the radio jet. 

We fitted the H$_2$ velocity field for the narrow component by an analytical model, assuming that the gas has circular
orbits in a plane of the galaxy, with the line-of-sight velocity given by \citep{bertola91}: 

\[
V_{\rm mod}(R,\Psi)=V_{s}+ \]

\begin{equation}
~~~~~\frac{AR\cos(\Psi-\Psi_{0})\sin(i)\cos^{p}(i)}{\left\{R^2[\sin^2(\Psi-\Psi_{0})+\cos^2(i)\cos^2(\Psi-\Psi_{0})]+C_o^2\cos^2(i)\right\}^{p/2}},
\label{model-bertola}
\end{equation}
where {\it R} is the projected distance to the nucleus with the corresponding position angle $\Psi$, $V_{s}$ is the galaxy's systemic velocity, {\it A} is the velocity amplitude, {\it i} is the disc inclination in relation to the plane of the sky ($i=0$ for a face on disc and $i=90^\circ$ for an edge on disc), 
$\Psi_0$ is the position angle of the line of nodes, {\it p} measures the slope of the rotation curve,  and $C_0$ is a concentration parameter. We used the {\sc mpfitfun} routine \citep{markwardt09} to perform the non-linear least-squares fit.

\begin{figure*}
\includegraphics[width=\textwidth]{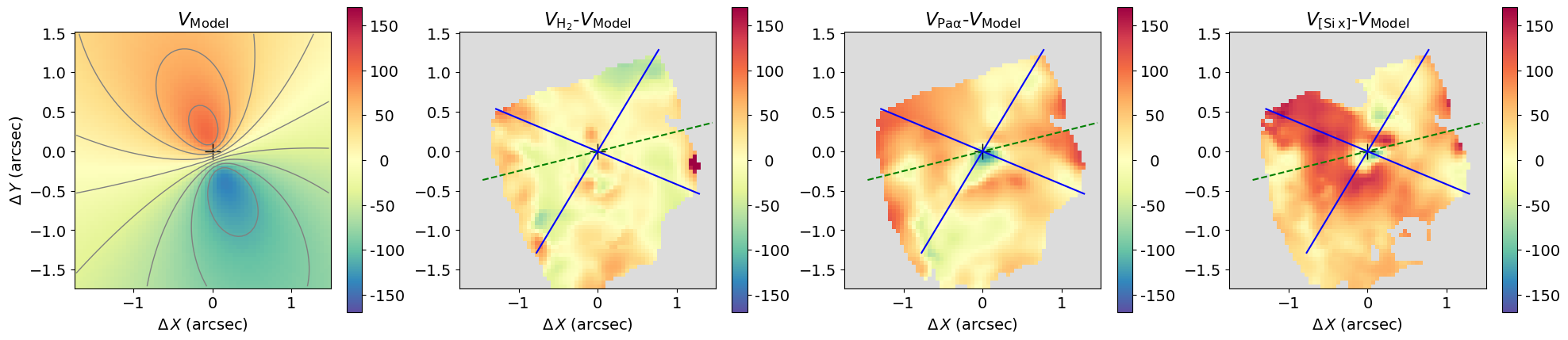}
\caption{From left to right:  Rotating disc model in the left panel and residual velocity maps (observed  -- model) for the H$_2$2.1218\,$\mu$m, Pa$\alpha$ and  [Si\,{\sc x}]1.4305\,$\mu$m emission lines. The dashed-green line shows the orientation of the radio jet and the blue lines delineate the bi-conical emission structure seen in the ionised gas flux distributions. }
\label{fig:residuals}
\end{figure*}

Figure~\ref{fig:residuals} show the rotating disc model in the left panel and the residual maps (observed velocities -- model) for the H$_2$2.1218\,$\mu$m, Pa$\alpha$ and  [Si\,{\sc x}]1.4305\,$\mu$m emission lines. For the H$_2$, the residuals are small in most locations, with a mean absolute value of $\sim$10 km\,s$^{-1}$ at distances closer than 1 arcsec from the nucleus. Some redshifts of $\sim$50 km\,s$^{-1}$ are seen  mainly to the southeast, at distances larger than $\sim$0.8 arcsec from the nucleus and within the ionisation cone. The residual maps for the ionised gas show redshifts, mainly within the ionisation cone, with velocities of up to $\gtrsim$100 km\,s$^{-1}$ in the southeast cone. These residuals are likely produced by outflows within the ionisation cone, as discussed in the next section.

The resulting best fit parameters are  $V_{s}=16795\pm7$ km\,s$^{-1}$, $A=195\pm14$ km\,s$^{-1}$, $p=1.44\pm0.06$, $\Psi_0=21^\circ\pm2^\circ$,  $C_0=0.23\pm0.05$\,arcsec, $i=30^\circ\pm7$ and the kinematical centre is consistent with the location of the continuum peak. 
If the accretion onto the black hole occurs along a preferred plane, the radio jet is expected to be observed perpendicularly to the inner part of the accretion disc. However,  
the comparison between the position angles of the jets and those of the dust disc major axes, on scales from tens to hundreds of parsecs,  shows that they are not aligned preferentially perpendicular to each other in radio galaxies \citep[e.g.][]{Schmitt02}. For Cygnus A, we find that the orientation of the line of nodes $\Psi_0$ is approximately perpendicular to the orientation of the radio jet, projected in the plane of the sky \citep[PA=104$^\circ$][]{Carilli96}.

\subsubsection{The multiphase outflows}

Previous near-IR long-slit spectroscopic observations of the central region of Cygnus A revealed the presence of blueshifted wings in the emission line profiles from the ionised gas, consistent with outflows in the northwest cone \citep[e.g.][]{Tadhunter03}. The blueshifted component is clearly detected in our NIFS spectra (Fig.~\ref{fig:fits}), its emission is spatially resolved, and we can further investigate the geometry of the outflows. For the ionised gas, we find that the blueshifted component emission extends from the nucleus to the northwest, up to the borders of the field of view (Fig.~\ref{fig:2GPaa}), while the H$_2$ emission lines show blueshifted wings only in the nucleus in  (Fig. \ref{fig:2GH2}).

The northwest is the near side of the bicone and the blueshifted component can be attributed to outflows within the northwestern cone. Assuming that the bicone axis and radio jet have the same orientation relative to the line of sight, making an angle of 30$^\circ$ relative to the plane of the sky \citep{Steenbrugge08} and a cone half-opening angle of 45$^{\circ}$, directly measured from the ionised gas emission-line flux maps (indicated by the blue lines in Figs.~\ref{fig:gh}, \ref{fig:2GPaa}, \ref{fig:2GH2} and \ref{fig:residuals}), the front wall of the northwestern cone makes an angle of 75$^\circ$ relative to the plane of the sky (15$^\circ$ relative the line-of-sight). The back wall of the northwestern cone is displaced by only 15$^\circ$, beyond the plane of the sky.  Similarly, the front wall of the southeastern cone is in front the plane of the sky, making an angle of 15$^\circ$ and the back wall of the cone is beyond the plane of the sky and makes an angle of 75$^\circ$ relative it. 

The blue wings detected in the emission lines to the northwest seem to be originated from outflowing gas located close to the front wall of the northwestern cone, with line-of-sight  velocities of up to 600\,km\,s$^{-1}$. The residual velocity maps between the velocity field of the narrow component and the rotating disc model (Fig.~\ref{fig:residuals}) for the ionised gas show redshifts to the southeast, with projected velocities of up to $\sim$150 km\,s$^{-1}$ inside the ionisation cone. These redshifts are likely produced by outflows inside the southeastern cone. We do not detect the highest velocity components, from the back wall of the southeastern cone, likely because it is obscured by dust, as it is beyond the  disc. Some residual redshifts are also seen to the northwest, which may be due to outflows closer to the back wall of the northwestern cone, which is 15$^\circ$, beyond the plane of the sky. Thus, the kinematics of the ionised gas in the inner 3.5$\times$3.5 kpc$^2$ of Cygnus A is consistent with a rotating disc component, as discussed in previous section, plus a bipolar outflow within the ionisation bicone. The blueshifts detected in H$_2$ could be originated by a compact molecular outflow produced by the interaction of the ionised outflows with the gas in the disc or by a lateral expansion of the dusty torus. Alternatively, the H$_2$ gas could be tracing the outer side of wide-angle winds, as suggested by theoretical studies and numerical simulations of the AGN driven winds \citep[e.g.,][]{silk98,ramos-almeida17,giustini19} and observed for nearby luminous Seyfert galaxies \citep{may17,may20}, where the molecular outflows are seen outside the borders of the ionisation cones.

We can use the observed kinematics and geometry of the outflows to roughly estimate the mass outflow rate  ($\dot{M}_{\rm out}$), by 
\begin{equation}
 \dot{M}_{\rm out} = 2\,m_p N_e v_{\rm out} f A,
\end{equation}
where $m_p$ is the mass of the proton, $N_e$ the electron density, $v_{\rm out}$ is the outflow velocity, $f$ is the filling factor, $A$ is the area of the cross section. We consider a circular cross section at 0.5 arcsec from the nucleus, with a radius of $\sim$600 pc for a half-opening cone angle of 45$^\circ$,  $f=1\times10^{-3}$ \citep[a typical value for the NLR;][]{martins03}, $N_{\rm e}=7900\,{\rm cm^{-3}}$ (the mean value derived in Sec.~\ref{sec:Ne}), and  $v_{\rm out}=600\,{\rm km\,s^{-1}}/\sin(75^\circ)\approx620\,{\rm km\,s^{-1}}$, where 600\,${\rm km\,s^{-1}}$ is the velocity of the blueshifted component (Fig.~\ref{fig:2GPaa}) and $75^\circ$ is the angle that the front wall of the northwestern cone makes with the plane of the sky, as discussed above. Under these assumptions and considering a filled cone, we obtain $\dot{M}_{\rm out}\approx280\, {\rm M_\odot\, yr^{-1}}$. The ionised gas mass outflow rates are usually highly uncertain, considering the number of assumptions that need to be made, regarding the geometry, filling factor and gas density \citep[e.g.][]{davies20}. For instance, if we assume $N_e\approx4450{\, \rm cm^{-3}}$ derived from the [\ion{Ar}{iv}]$\lambda\lambda$4711,4740 (Sec. \ref{sec:Ne}), we obtain $\dot{M}_{\rm out}\approx160\, {\rm M_\odot\, yr^{-1}}$. Further, if we assume that the geometry of the outflows is a hollow bicone, as suggested for nearby Seyfert galaxies \citep[e.g.][]{sb_N4151_kin,fischer13}, we find $\dot{M}_{\rm out}\approx100 \, {\rm M_\odot\, yr^{-1}}$ using inner and outer half-opening angles of $40^\circ$ and $50^\circ$, respectively. Using the density estimated from the [S\,{\sc ii}] lines in Sec.~\ref{sec:Ne}, the resulting outflow rate would be about one order of magnitude smaller, but as pointed out in \citet{davies20}, the electron density derived from the [S\,{\sc ii}] lines are significantly smaller that the real density of the outflows. Thus, the mass outflow-rate in ionised gas for Cygnus A may be in the range $\sim100-280\, {\rm M_\odot\, yr^{-1}}$. 
The AGN in Cygnus A has a bolometric luminosity of (0.5--2.0)$\times$10$^{46}$ erg s$^{-1}$ \citep{Tadhunter03} and the derived $\dot{M}_{\rm out}$ values above is similar to the ionised outflow rates observed for AGN with similar luminosities \citep[e.g.][]{davies20,kakkad20,vayner21,bruno21}

We can use the above range of mass outflow rate to estimate the  kinetic power of the outflow by
\begin{equation}
\dot{E}\approx\frac{\dot {M}_{out}}{2}(v_{\rm out}^2+3\sigma_{\rm out}^2),
\end{equation}
where $\sigma_{\rm out}$ is the velocity dispersion of the outflow.  Using $\sigma\approx700\,$km\,s$^{-1}$ (from Fig.\,\ref{fig:2GPaa}) we obtain 
$\dot{E}\approx(6-16)\times10^{43}$ erg\,s$^{-1}$, which corresponds to 0.3--3.3 per cent of the AGN bolometric luminosity. If the kinetic power of the outflow is above 0.5 per cent of the AGN luminosity it becomes effective in suppressing the star formation in the host galaxy \citep{hopkins_elvis10}. Thus, the ionised outflows in Cygnus A seem to be powerful enough to affect the evolution of the galaxy. 
Molecular gas phases of the outflow may also be present, as indicated by the hot H$_2$ blueshifted gas. In addition,  numerical simulations indicate that the kinetic energy corresponds to less than 20 per cent of the  energy of the total outflow \citep{richings18b} and thus, the power of the multi-phase outflow in Cygnus A may be even larger than the value derived above.

\section{Conclusions}
\label{concsec}

We have analyzed H, K$_s$ and K-band spectra from the inner  3.5$\times$3.5 kpc$^2$ of the radio galaxy Cygnus A, obtained with the Gemini NIFS at a spatial resolution of $\approx$200~pc and velocity resolution of $\approx$\,50\,km\,s$^{-1}$. We have studied the excitation and kinematic properties of the molecular and ionised emitting gas. The main conclusions of this work are:
 
 \begin{itemize}
     
\item The emission-line flux distributions for the ionised gas ([Si\,{\sc x}]1.4305\,$\mu$m, Pa$\alpha$, [S\,{\sc xi}]1.9201\,$\mu$m and
[Si\,{\sc vi}]1.9630\,$\mu$m) clearly show a biconical morphology, with the bicone axis observed along the orientation of the radio jet and half-opening angle of 45$^\circ$. The [Fe\,{\sc ii]} emission lines show a more round flux distribution than those of lines from other ionised species and the H$_2$ emission is observed mostly in regions outside the ionisation cone.

\item  Coronal line emission is observed up to the borders of the NIFS field of view ($\sim$1.5 arcsec/1.75 kpc). Higher ionisation gas is seen in the easterly cone than in the  westerly cone, as observed in the [Si\,{\sc vi}]/Pa$\alpha$ and [Si\,{\sc x}]/Pa$\alpha$ intensity ratio maps.  

\item Emission-line ratio diagnostic diagrams indicate that the H$_2$ and [Fe\,{\sc ii}] emission lines are consistent with excitation by the central AGN, with some contribution of shocks in a region extending from the nucleus to southwest.

\item From the Pa$\alpha$/Br$\gamma$ line ratio we derive visual extinctions of up to 20 mag and an average value of 12.5 mag.  The average value of the electron density is $\sim$7900$\,{\rm cm^{-3}}$ as obtained from the  [Fe\,{\sc ii}]1.5330\,$\mu$m/[Fe\,{\sc ii}]1.6440\,$\mu$m emission-line ratio. Both density and visual extinction are larger than the values derived using optical lines, consistent with the fact that near-IR observations can penetrate more deeply into the gas emission structure and probe denser and more obscured regions.

\item We observe two gas kinematic components produced by (i) a rotating disc with major axis oriented along $\Psi_0=21^\circ\pm2^\circ$ and projected velocity amplitude of $\sim$150\,km\,s$^{-1}$, clearly observed in the H$_2$ velocity field and (ii) outflows within the bicone, with velocities of up to 600 km\,s$^{-1}$ observed in ionised gas. Compact nuclear H$_2$ outflows are seen in the inner 0.5 arcsec, produced by the interaction of the ionised outflows with the gas in the disc, by a lateral expansion of the dusty torus or by the outer side of the biconical winds seen in ionised gas.

\item The geometry of the outflows is consistent with a bicone with axis along the orientation of the radio jet, half-opening angle of 45$^\circ$, making an angle of 30$^\circ$ relative to the plane of the sky with the  northwest being the near side of the bicone. Using this geometry along with the observed kinematics, we obtain ionised mass outflow rates in the range $\sim100-280 {\rm M_\odot\, yr^{-1}}$. 

\item The kinetic power of the outflows corresponds to 0.3--3.3 per cent of the bolometric luminosity of the AGN in Cygnus A, being powerful enough to
suppress star formation in the host galaxy.

 \end{itemize}

\section*{Acknowledgments}
The author thanks an anonymous referee for valuable comments that helped to improve this paper.
This study was financed in part by Conselho Nacional de Desenvolvimento Cient\'ifico e Tecnol\'ogico (202582/2018-3, 304927/2017-1 and 400352/2016-8) and Funda\c c\~ao de Amparo \`a pesquisa do Estado do Rio Grande do Sul (17/2551-0001144-9 and 16/2551-0000251-7). Based on observations obtained at the Gemini Observatory, which is operated by the Association of Universities for Research in Astronomy, Inc., under a cooperative agreement with the NSF on behalf of the Gemini partnership: the National Science Foundation (United States), National Research Council (Canada), CONICYT (Chile), Ministerio de Ciencia, Tecnolog\'{i}a e Innovaci\'{o}n Productiva (Argentina), Minist\'{e}rio da Ci\^{e}ncia, Tecnologia e Inova\c{c}\~{a}o (Brazil), and Korea Astronomy and Space Science Institute (Republic of Korea). 
This research has made use of NASA's Astrophysics Data System Bibliographic Services. 
This research has made use of the NASA/IPAC Extragalactic Database (NED), which is operated by the Jet Propulsion Laboratory, California Institute of Technology, under contract with the National Aeronautics and Space Administration. 
\section*{Data Availability}
The data used in this paper is available in the Gemini Science Archive at https://archive.gemini.edu/searchform under the project code GN-2006A-C-11. Processed datacubes used will be shared on reasonable request to the corresponding author.



\bibliographystyle{mnras}
\bibliography{paper} 



\bsp	
\label{lastpage}
\end{document}